\documentclass[english,pra,superscriptaddress,showpacs,twocolumn]{revtex4}
\usepackage[T1]{fontenc}
\usepackage[latin9]{inputenc}
\usepackage{array}
\usepackage{mathrsfs}
\usepackage{amsmath}
\usepackage{amssymb}
\usepackage{graphicx}

\makeatletter

\providecommand{\tabularnewline}{\\}

\@ifundefined{textcolor}{}
{%
 \definecolor{BLACK}{gray}{0}
 \definecolor{WHITE}{gray}{1}
 \definecolor{RED}{rgb}{1,0,0}
 \definecolor{GREEN}{rgb}{0,1,0}
 \definecolor{BLUE}{rgb}{0,0,1}
 \definecolor{CYAN}{cmyk}{1,0,0,0}
 \definecolor{MAGENTA}{cmyk}{0,1,0,0}
 \definecolor{YELLOW}{cmyk}{0,0,1,0}
}

\usepackage{times}
\usepackage{color}
\usepackage{amsmath}
\usepackage{graphicx}
\usepackage{amssymb}
\usepackage{esint}

\makeatother

\usepackage{babel}
\begin{document}

\title{Excitation spectrum for an inhomogeneously dipole-field-coupled superconducting
qubit chain}

\author{Hou Ian}

\affiliation{Advanced Science Institute, RIKEN, Wako-shi, Saitama 351-0198, Japan}

\affiliation{Faculty of Science and Technology, University of Macau, Macau}

\author{Yu-xi Liu}

\affiliation{Advanced Science Institute, RIKEN, Wako-shi, Saitama 351-0198, Japan}

\affiliation{Institute of Microelectronics, Tsinghua University, Beijing 100084,
China}

\affiliation{Tsinghua National Laboratory for Information Science and Technology
(TNList), Tsinghua University, Beijing 100084, China}

\author{Franco Nori}

\affiliation{Advanced Science Institute, RIKEN, Wako-shi, Saitama 351-0198, Japan}

\affiliation{Physics Department, The University of Michigan, Ann Arbor, MI 48109,
USA}
\begin{abstract}
When a chain of $N$ superconducting qubits couples to a coplanar
resonator in a cavity, each of the qubits experiences a different
dipole-field coupling strength due to the waveform of the cavity field.
We find that this inhomogeneous coupling leads to a dependence of
the qubit chain's ladder operators on the qubit-interspacing $l$.
Varying the spacing $l$ changes the transition amplitudes between
the angular momentum levels. We derive an exact diagonalization of
the general $N$-qubit Hamiltonian and, through the $N=4$ case, demonstrate
how the $l$-dependent operators lead to a denser one-excitation spectrum
and a probability redistribution of the eigenstates. Moreover, we
show that the variation of $l$ between its two limiting values coincides
with the crossover between Frenkel- and Wannier-type excitons in the
superconducting qubit chain. 
\end{abstract}


\maketitle

\section{Introduction}

Superconducting quantum circuits have attracted considerable attention
because of their capabilities (i) to demonstrate macroscopically the
basic interaction of one ``atom'' and one photon in a cavity (e.g.,~\cite{jqyou03,blais04}),
(ii) to serve as a platfrom for testing many quantum optical phenomena
(e.g.,~\cite{hauss08,ian10,shevchenko10,jqyou11,wilson11}), as well
as (iii) to show its potential as a basis for quantum information
processing (e.g.,~\cite{jqyou05,yxliu06,hemler09,buluta11}).

While research on single-qubit interactions are more common, many
recent articles also studied multi-qubit interactions. Superconducting
circuits with such interactions are also known as quantum metamaterials~\cite{rakhmanov08}.
To be precise, the circuit system we consider here consists of a chain
of $N$ superconducting two-level qubits coupled to a photon mode
in a superconducting coplanar waveguide resonator. Compared to the
single-qubit version, it can manifest even more quantum phenomena,
including plasma waves~\cite{savelev10}, controllable collective
dressed states~\cite{fink09}, and quantum phase transitions~\cite{wang07,lambert09,nataf10,tian10}.
It also promises potential for various applications, including quantum
simulators~\cite{buluta09} and quantum memories~\cite{diniz11}.

Theoretical studies of multi-qubit interactions with a photon often
employ the Dicke model~\cite{dicke54}, where the Pauli operators
are summed and transformed into a bosonic operator. In this approach,
the chain of qubits is treated collectively as an atomic ensemble
and the excited qubits are collectively regarded as one exciton mode.
This theoretical simplification proves adequate when (i) the number
of excitations in the system is low (in the so-called ``one-photon''
processes) and (ii) the number of qubits is large enough such that
the interspacing $L_{\mathrm{q}}$ between neighboring qubits can
be ignored compared to the photon wavelength $L_{\mathrm{p}}$ in
the resonator (i.e., the qubits can be regarded as a continuum).

However, the question of how excitations arise in superconducting
metamaterials when these two conditions are \emph{not} met remains
unanswered. In a realistic setting for a superconducting circuit,
the number $N$ of qubits present can range from one to, say, 10,
but $N$ would not be as large as the number of atoms we usually have
for an alkaline atomic ensemble in an optical microcavity, which is
typically greater than $10^{5}$. Therefore, the Dicke model, which
treats $N\to\infty$, does \emph{not} apply well to the case of multi-qubit
superconducting circuits with $N\leq10$.

When a chain of superconducting circuit qubits is arranged as a one-dimensional
array (i.e., a superconducting qubit chain or SQC), each qubit is
\emph{inhomogeneously coupled} to the circuit photon mode. In other
words, each qubit has a \emph{different} coupling strength to the
traversing photon field. This occurs naturally since, unlike its optical
cavity QED counterpart, the photon wavelength $L_{\mathrm{p}}$ is
comparable to the qubit interspacing $L_{\mathrm{q}}$ in a superconducting
circuit. The coupling strength thus depends on the position of the
qubit relative to the photon waveform. The effect of the varying coupling
strength becomes even more obvious if multi-mode couplings are taken
into consideration. For example, the qubits on the antinodes of the
waveform will couple most strongly, whereas those on the nodes will
not couple.

The first step to understand and characterize this\emph{ }inhomogeneously-coupled
system (the aim of this article) is to obtain the energy spectrum
of the collective excitation mode in the chain of qubits and to compare
it with that of the Dicke model. We find that the inhomogeneity of
the couplings incurs an algebraic deformation of the Pauli operators
of the qubits~\cite{polychron90,rocek91,delbecq93,bonatsos93}. We
quantify this deformation through a ``deformation factor,'' which
is a function of the relative spacing $l=2L_{\mathrm{q}}/L_{\mathrm{p}}$,
and characterize the amount the inhomogeneous system deviates from
the homogeneous case. The deformation factor modifies the spin operators
of the collective qubit chain. Consequently, the excitation spectrum
will not only be a function of the eigenenergy of the photon mode
and the qubit level spacing, but is also highly related to the deformation
factor and hence the relative spacing $l$.

Note that when atoms are confined to a cavity, the magnetic or laser
field that is exerted on them is \emph{uniform}. The strength of the
interaction can be uniformly increased or decreased according to the
density of the atoms. This macroscopic viewpoint does not differentiate
between the identities of the atoms. However, for circuit QED, the
identities of the qubits are partially differentiated since the qubits
can be categorized according to the values of their coupling strength
to the photon mode. This partial differentiation has made understanding
the inhomogeneous system a many-body physics question.

Our deformation algebraic approach here is a statistical approximation
method that can be regarded as finding the average contribution of
the coupling strength given by the SQC as a whole. In the end, the
characterization (the excitation spectrum) of the SQC as an inhomogeneous
system is not parametrized by the individual qubits, but by the relative
spacing $l$. In other words, the spacing $l$ is one extra degree
of freedom peculiar to the inhomogeneous SQC, not seen in a homogeneous
optical cavity.

We will first introduce the model and derive the deformation factor
in Sec.~\ref{sec:model}. With the deformation factor, new operation
rules for the spin angular momentum operators are found by solving
a difference equation in Sec.~\ref{sec:op_rules}. The general energy
spectrum for $n$-qubit SQC is given in Sec.~\ref{sec:spectrum}.
We also derive in Sec.~\ref{sec:spectrum} a one-excitation spectrum
for a 4-qubit SQC as a nontrivial case to show the effects of the
inhomogeneity. Namely, the energy splittings between the eigenstates
of the deformed coupling case shrink, while the probability amplitudes
of the eigenstates are redistributed such that higher-photon occupations
are favored.

In the final Sec.~\ref{sec:crossover}, we will consider how the
collective excitations on the SQC would emulate the excitons in atomic
lattices. In one limit, it becomes a Wannier-type exciton~\cite{Mahan},
where the wave function of the excited level is localized on a single
atom. In the opposite limit, it emulates a Frenkel-type exciton, which
has an extended wave function across multiple atoms. Changing the
degree of freedom $l$ lets the emulated exciton undergo a crossover
between these two types of excitons. This crossover depends on a $(2N-1)$-th
order trigonometric equation, whose solution corresponds to the asymptotic
turning point from the deformed (inhomogeneously-coupled) SQC to the
undeformed (homogeneously-coupled) SQC.

\section{Inhomogeneous coupling model\label{sec:model}}

\subsection{Inhomogeneous coupling}

For a finite number $N$ of spins in the SQC, the problem discussed
here is similar to the Tavis-Cummings (TC) model~\cite{tavis68,swain72},
where all the spins are grouped into a total ``large'' spin. However,
the exactly solvable TC model applies only when the coupling is homogeneous
and when the eigenfrequencies between the qubits and the photon mode
are equal. When the coupling is inhomogeneous, the large spin does
not obey the usual commutation relations of the Pauli matrices, which
the TC-model assumes.

The new commutation relations of the large spin introduced by the
inhomogeneity are pertinent to the deformed SU(2) Lie algebras. From
these algebraic structures, we can establish a deformed dipole-field
coupling model, of which the TC model is a special case. In the following
discussion, we consider the typical case where the inter-qubit spacing
is uniform. The coupling strength of each qubit to the photon field
can, therefore, be written as a cosine function of a phase factor
which is determined by the position $jl$ of the $j$-th qubit, where
$j$ is the reduced coordinate and $l$ is the relative spacing introduced
above.

\begin{figure}
\includegraphics[bb=75bp 40bp 775bp 330bp,clip,width=8.1cm]{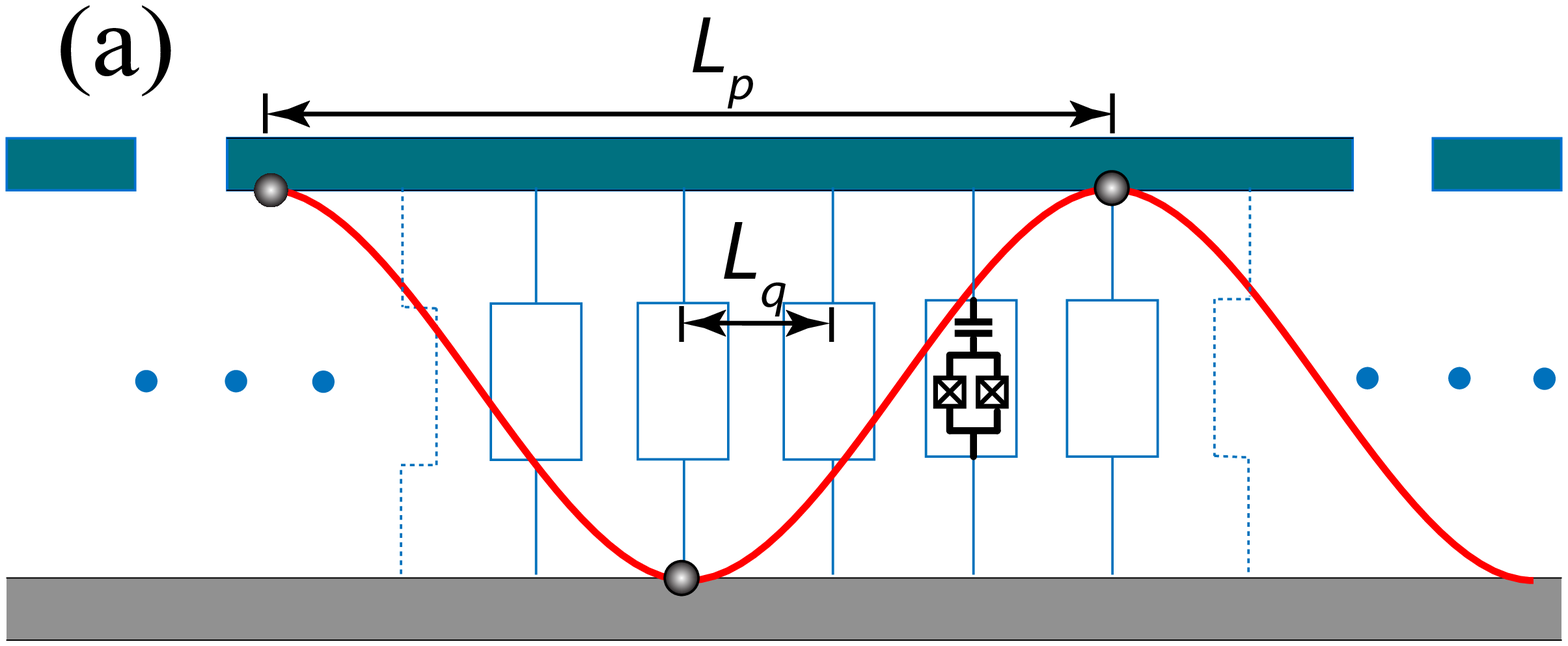}

\includegraphics[bb=120bp 192bp 490bp 585bp,clip,width=5.6cm]{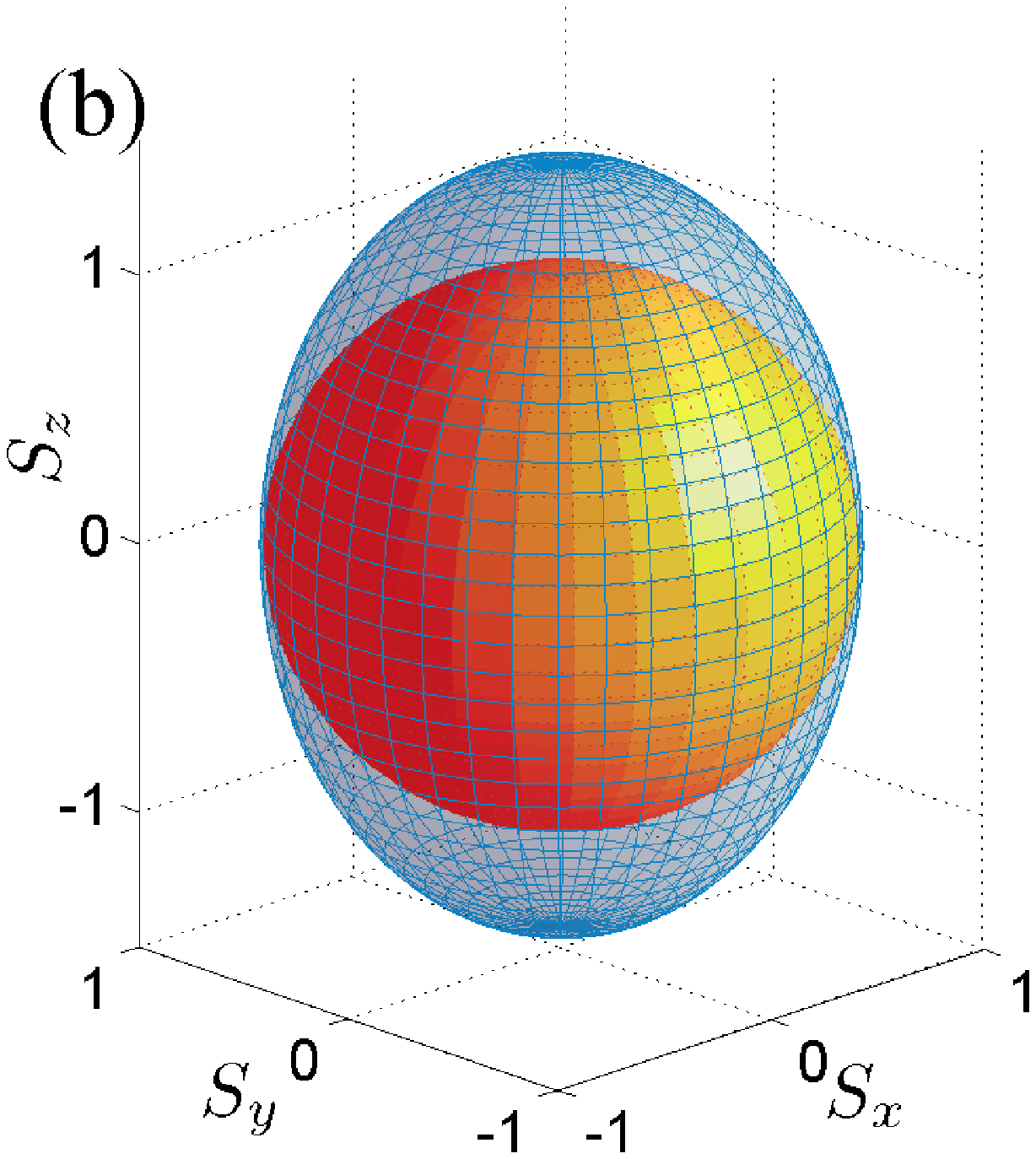}

\caption{(Color online) (a) Schematic diagram of an SQC with spacing $l=1/3$.
The upper strip represents a coplanar resonator whose potential waveform
of wavelength $L_{\mathrm{p}}$ is drawn as a sinusoidal curve, while
the lower one represents a ground strip. The rectangles between the
strips are qubits with interspacing $L_{\mathrm{q}}$. The gray dots
on the curve indicate the anti-nodes of the waveform. (b) The elongation
effect on the large spin due to inhomogeneous coupling. The ellipsoid
shows a case of deformation $R_{N,l}=0.5$. The unit sphere shows
the spin under the usual homogeneous coupling.\label{fig:model}}
\end{figure}

The situation is illustrated in Fig.~\ref{fig:model}(a). A chain
of qubits is sandwiched between the superconducting coplanar resonator
and a superconducting ground strip. The photon mode providing different
potential energies on the spins is shown by the red sinusoidal curve.

We use the operators $\{\sigma_{j,z}\}$ to denote Josephson junction
qubits, and $\{a,a^{\dagger}\}$ to denote the operators for the single-photon
mode. With the wave vector being the reciprocal of the photon field
wavelength on the one-dimensional lattice, $k=2\pi/L_{\mathrm{p}}$,
the dipole-field coupling is of the form $\sigma_{j,x}(a\cos(j\pi l)+\mathrm{h.c.})$.
Under the rotating wave approximation, the Hamiltonian can be written
as ($\hbar=c=1$) 
\begin{equation}
H=\omega_{\mathrm{q}}\sum_{j=0}^{N-1}\sigma_{j,z}+\omega_{0}a^{\dagger}a+\eta\sum_{j=0}^{N-1}\cos(j\pi l)\left[\sigma_{j,+}a+\sigma_{j,-}a^{\dagger}\right]\label{eq:ham}
\end{equation}
where $\omega_{\mathrm{q}}$ is the eigenenergy of the spins, $\omega_{0}$
the mode frequency of the photon, and $\eta$ the coupling amplitude.

To diagonalize the Hamiltonian in Eq.~\eqref{eq:ham}, we introduce
the ``large spin'' operators: the magnetic moment or $z$-direction
collective spin operator 
\begin{equation}
S_{z}=\sum_{j=0}^{N-1}\sigma_{j,z},
\end{equation}
which is no different from the homogeneous case, and the paired raising
and lowering operators 
\begin{eqnarray}
S_{+} & = & \sum_{j=0}^{N-1}\sigma_{j,+}\cos(j\pi l)\\
S_{-} & = & \sum_{j=0}^{N-1}\sigma_{j,-}\cos(j\pi l)
\end{eqnarray}
which have the special sinusoidal dependence on $l$ due to the inhomogeneity.
The commutator of the paired ladder operators \emph{no} longer equals
to $2S_{z}$ but has an additional term due to the cosine coefficients,
i.e., $[S_{+},S_{-}]=2\Sigma_{z}$ with

\begin{equation}
\Sigma_{z}=S_{z}+\sum_{j=0}^{N-1}\sin[j\pi(1+l)]\sin[j\pi(1-l)]\sigma_{j,z}.\label{eq:Sigma_z}
\end{equation}
The detailed derivation is shown in Appendix~\ref{sub:comm_rel}.
Note that in the usual circuit QED system~\cite{blais04}, where
only one spin is placed at midway, the spacing $l$ equals to 2, for
which the latter term in $\Sigma_{z}$ vanishes. This is the limiting
case which corresponds to the Wannier type of excitation, where the
set of spin operators retains the usual structure of an undeformed
SU(2) algebra.

\subsection{Deformed algebraic structure}

When the second term of $\Sigma_{z}$ does not vanish, the algebraic
structure is called deformed~\cite{polychron90,rocek91,delbecq93,bonatsos93}.
In order to quantify the deformation, the commutator of the ladder
operators needs to be expressed as a function of $S_{z}$, i.e., $\Sigma_{z}=f(S_{z})$.
To find this function $f$, we consider an underlying manifold, on
which there is a local point, say the origin 0, where we define a
tangent space with the Pauli $z$-matrices $\{\sigma_{j,z}\}$ being
its basis vectors, since these matrices are linearly independent.
The operator $S_{z}$, defined above with uniform coefficients, can
be deemed a vector in this tangent space; the operator $\Sigma_{z}$
is then another vector dependent on the parameter $l$ and is a deviation
or deformation from $S_{z}$. Thus, the first-order approximation
of $\Sigma_{z}$ with respect to $S_{z}$ is its projection onto the
vector $S_{z}$. That is, since the cosine coefficients are bounded,
we can use their Hilbert-Schmidt norm 
\begin{equation}
\left\langle \Sigma_{z},S_{z}\right\rangle =\mathrm{tr}\left(\Sigma_{z}^{\ast}S_{z}\right)=N+\sum_{j=0}^{N-1}\cos(2j\pi l)\label{eq:HS_norm}
\end{equation}
and the Schmidt decomposition~\cite{reed_simon} to write $\Sigma_{z}=R_{N,l}S_{z}$
as a deformation of the original $z$-spin operator where 
\begin{equation}
R_{N,l}=\frac{1}{4N}\left\{ 2N+1+\frac{\sin[(2N-1)\pi l]}{\sin(\pi l)}\right\} \label{eq:defm}
\end{equation}
is the deformation factor (Cf. Appendix~\ref{sub:defm_factor} for
this derivation). The commutator of the ladder operators can now be
expressed as 
\begin{equation}
[S_{+},S_{-}]=2R_{N,l}S_{z},\label{eq:defm_comm_rel}
\end{equation}
where $R_{N,l}$ has a limiting value of one when $l\to0$ or $l\to\infty$,
for which the usual structure used in the TC model is retained.

Since the deformation factor $R_{N,l}$ does not affect the commutation
relations between the ladder operators and the $z$-spin, the large-spin
operators $\{S_{z},S_{+},S_{-}\}$ form a specific deformed algebra~\cite{polychron90,rocek91}
and not the more general type~\cite{delbecq93}. The Casimir operator
\begin{equation}
C=S_{-}S_{+}+h(S_{z})
\end{equation}
of the algebra, which equals to the undeformed spin momentum square,
$S^{2}=S_{x}^{2}+S_{y}^{2}+S_{z}^{2}$, for the homogeneous coupling
case, is accordingly deformed. Through solving a recursive relation
(Cf. Appendix~\ref{sub:csmir_op} for details), we find $h(S_{z})=R_{N,l}(S_{z}^{2}+S_{z})$
and hence
\begin{equation}
C=S_{x}^{2}+S_{y}^{2}+R_{N,l}S_{z}^{2},\label{eq:csmir_op}
\end{equation}
which shows that the spin momentum is \emph{reduced} along the $z$-direction:
\begin{equation}
S^{2}=S_{x}^{2}+S_{y}^{2}+R_{N,l}S_{z}^{2}.
\end{equation}
To visualize this reduction of the spin momentum, we can take a unit
value for the spin moment and let the Casimir operator be represented
by a unit sphere in 3-dimensional space for the undeformed case. With
$R_{N,l}\leq1$, the deformation would be an elongation of the unit
sphere, along the $z$-axis, to an ellipsoid, while the $x$- and
$y$-semi-minor axes remain unchanged, as shown in Fig.~\ref{fig:model}(b).

\section{Operation rules\label{sec:op_rules}}

If we consider the unit sphere of Fig.~\ref{fig:model}(b) as a Bloch
sphere on which the large spin prescribes its $N$ levels, its elongation
due to deformation will accordingly modify the transitions between
the levels. Since the spin-up and the spin-down momenta do not change,
which are still $\omega_{\mathrm{q}}N/2$ and $-\omega_{\mathrm{q}}N/2$,
the narrow part of the ellipsoid effectively squeezes the transition
probabilities.

More precisely, we consider an arbitrary eigenstate $\left|r,m\right\rangle $,
for which 
\begin{eqnarray}
S_{z}\left|r,m\right\rangle  & = & m\left|r,m\right\rangle \label{eq:S_z}\\
S^{2}\left|r,m\right\rangle  & = & r(r+1)\left|r,m\right\rangle .\label{eq:S^2}
\end{eqnarray}
The ladder operators result in an $(r,m)$-dependent off-diagonal
matrix element $\alpha_{m}^{(r)}$, i.e., 
\begin{eqnarray}
S_{+}\left|r,m\right\rangle  & = & \alpha_{m}^{(r)}\left|r,m+1\right\rangle \label{eq:S_+}\\
S_{-}\left|r,m\right\rangle  & = & \alpha_{m-1}^{(r)}\left|r,m-1\right\rangle \label{eq:S_-}
\end{eqnarray}
By examining the diagonal elements of the commutator of the ladder
operators, we find a difference equation $(\alpha_{m}^{(r)})^{2}-(\alpha_{m-1}^{(r)})^{2}=-2mR_{N,l}$.
With the value $\alpha_{-r}^{(r)}=0$, the equation can be solved
to give the deformed off-diagonal matrix elements or transition probabilities
for the ladder operators 
\begin{equation}
\alpha_{m}^{(r)}=\sqrt{R_{N,l}(r-m)(r+m+1)}.\label{eq:op_rule}
\end{equation}
See Appendix~\ref{sec:derive_op_rules} for its derivation.

Geometrically speaking, the deformation process is a homeomorphism
with a redefined metric $g=(1,1,R_{N,l})$. Since $R_{N,l}\leq1$,
the metric norm is less than unity. The deformation does not affect
the level spacings of the magnetic moment $S_{z}$: the number $m$
still takes $(2r+1)$ values (i.e., the ellipsoid is homeomorphic
to the sphere). But the transition amplitudes to traverse the sphere
decrease: if we start with a spin-up state $\left|r,r\right\rangle $
and finish with a spin-down state $\left|r,-r\right\rangle $, then
all iterations with $S_{-}\left|r,m\right\rangle =\alpha_{m-1}^{(r)}\left|r,m-1\right\rangle $
have $\alpha_{m}^{(r)}$ smaller than the original $\bar{\alpha}_{m}^{(r)}=\sqrt{(r-m)(r+m+1)}$
(i.e., the ellipsoid is not isometric to the sphere).

\begin{figure}
\includegraphics[bb=72bp 225bp 450bp 550bp,clip,width=6cm]{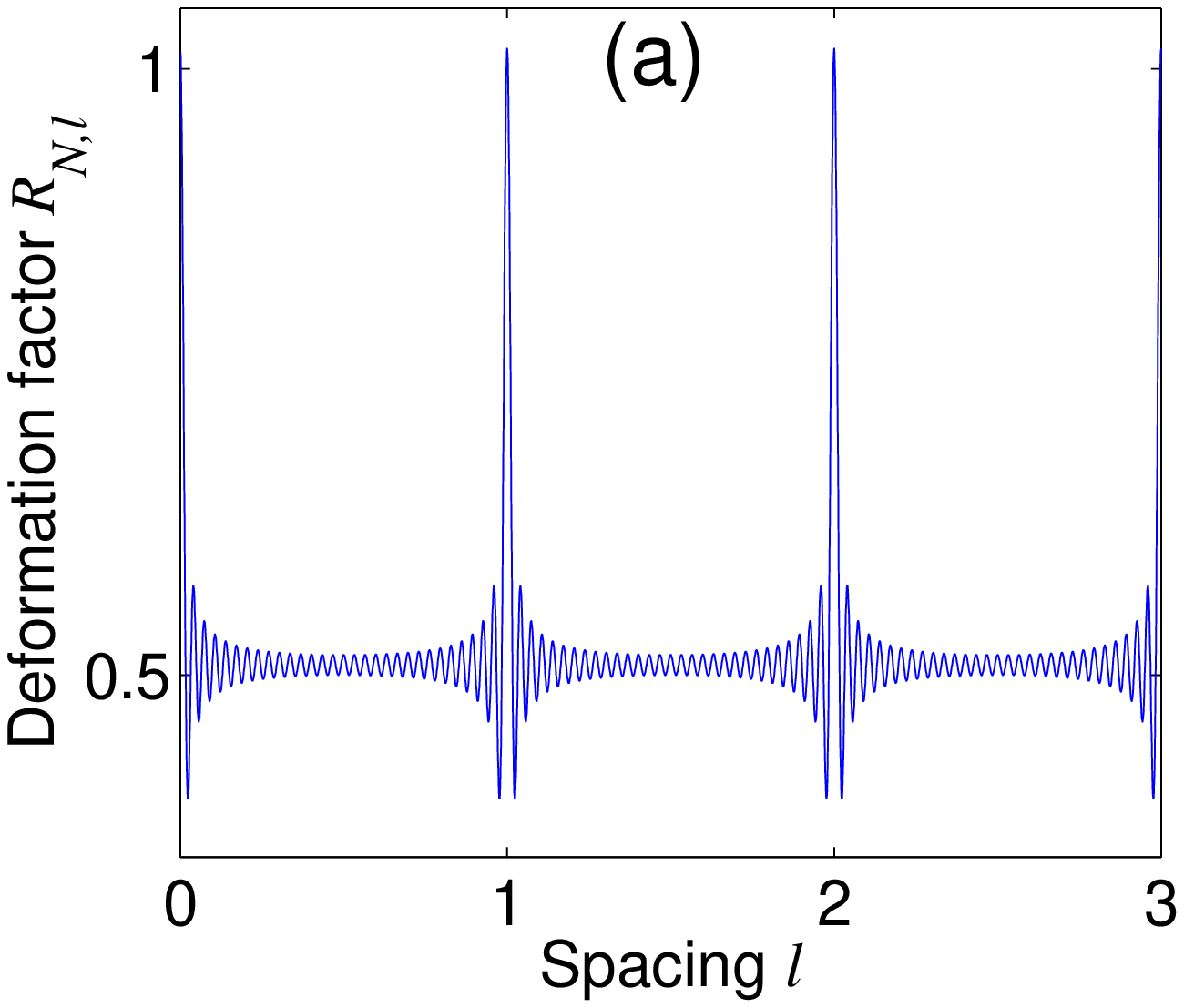}

\includegraphics[bb=98bp 190bp 480bp 610bp,clip,width=6cm]{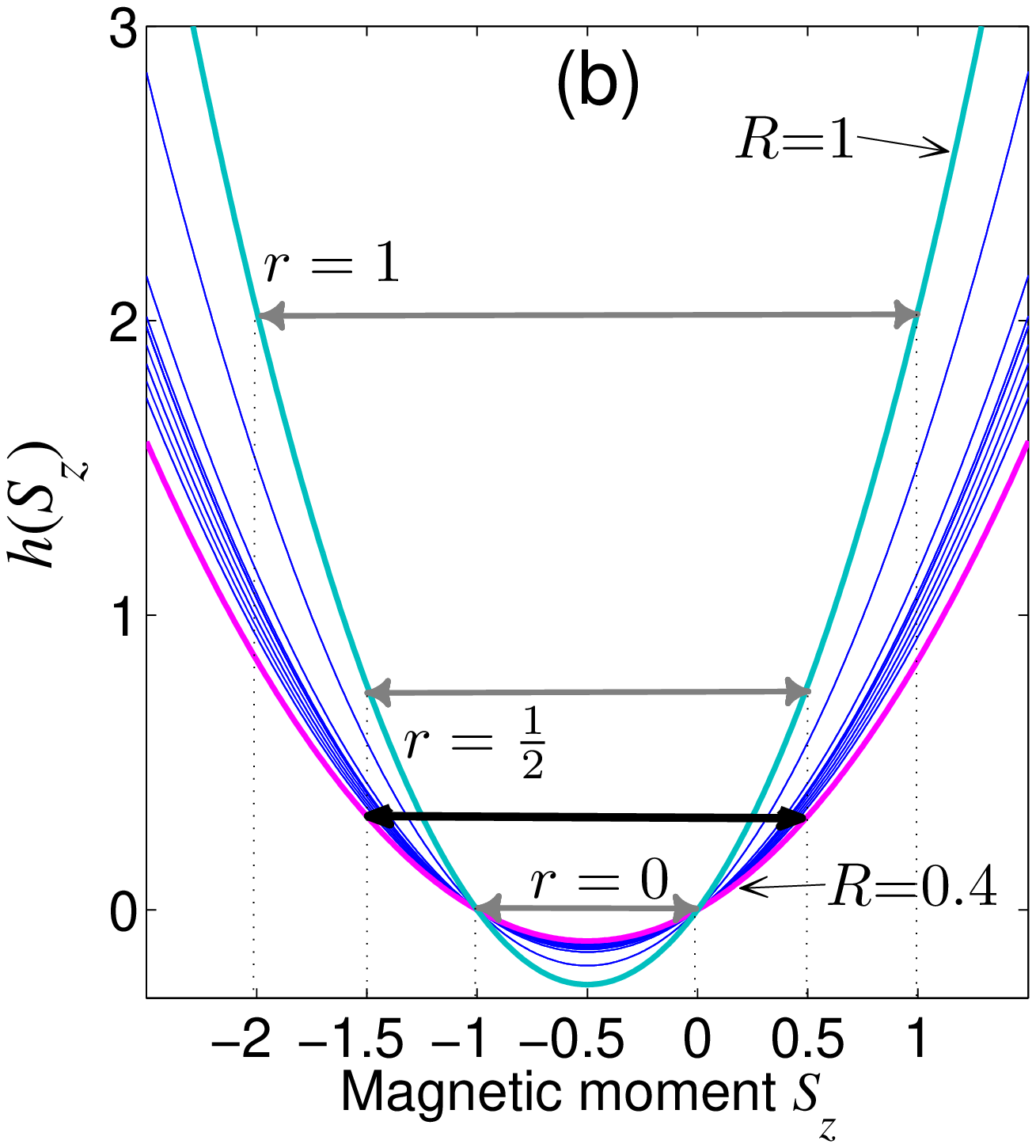}

\caption{(Color online) (a) The deformation factor $R_{N,l}|_{N=30}$, as an
oscillating function of the qubit spacing $l$. (b) Plot of the function
$h(S_{z})$ for various values of $R_{N,l}$. The top and bottom boldened
curves correspond to $R_{N,l}=1$ and $0.4$, respectively. The gray-shaded
double-arrows indicate three spin levels of the\emph{ }undeformed\emph{
}SQC, while the black arrows show the $r=1/2$ level of the deformed
case.\label{fig:deform_csmir}}
\end{figure}

The deformation factor expressed in Eq.~\eqref{eq:defm} is an oscillating
function of $l$, where the sine in the numerator determines the period
of oscillation and the sine in the denominator determines the period
of the envelope. Therefore, the spin angular momentum of the SQC would
be \emph{oscillating} between the unit sphere and the ellipsoid, depending
on the qubit spacing $l$. The plot of $R_{N,l}$ in Fig.~\ref{fig:deform_csmir}(a)
for an SQC of $N=30$ qubits shows a typical case with envelop of
period 1 and local minimum of $0.4$. The function $h(S_{z})$ associated
with this deformation factor is a parabola of the magnetic moment
$S_{z}$. For a nontrivial deformation $R_{N,l}<1$, this parabola
flattens and the spin levels become denser. The curvature of the parabola
decreases while its minimum value $-R_{N,l}/4$ increases. As shown
in Fig.~\ref{fig:deform_csmir}(b), the black (gray) arrow indicates
the spin level $r=1/2$ for the deformed $R_{N,l}=0.4$ (undeformed
$R_{N,l}=1$) case of SQC. So varying $l$ makes the curve $h(S_{z})$
oscillate between the boldened curves that correspond to $R_{N,l}=0.4$
and $R_{N,l}=1$, respectively. We can also observe that the level
splittings are reduced, reflecting the elongated structure of the
Bloch sphere in Fig.~\ref{fig:model}(b) and the modified operation
rule of Eq.~\eqref{eq:op_rule}.

\section{Deformed spectrum\label{sec:spectrum}}

Equipped with the modified operation rules, we can diagonalize the
Hamiltonian in Eq.~\eqref{eq:ham}. First, we split the Hamiltonian
into two parts
\begin{eqnarray}
H_{0} & = & \omega_{0}(S_{z}+a^{\dagger}a),\label{eq:H_0}\\
H_{1} & = & \tilde{\omega}_{0}a^{\dagger}a+\eta(S_{+}a+S_{-}a^{\dagger}).\label{eq:H_1}
\end{eqnarray}
Let $u$ be the number of total excitations and hence the eigenvalue
of $H_{0}$. Let $n$ be the number of photons in the system such
that the SQC magnetic moment is $m=u-n$. Let $\nu$ be the eigenvalue
of the interaction part $H_{1}$, where $\tilde{\omega}_{0}=\omega_{0}-\omega_{\mathrm{q}}$.

The eigenvector of the Hamiltonian can then be expanded as a superposition
of different configurations of photon number and spin states:
\begin{equation}
\left|u,r\right\rangle =\sum_{n}c_{n}\left|n;r,u-n\right\rangle .
\end{equation}
The expansion coefficients $c_{n}$ satisfy a recursive relation~\cite{swain72}:
\begin{equation}
c_{n+1}\sqrt{n+1}\alpha_{u-(n+1)}^{(r)}-c_{n}\tilde{v}_{n}+c_{n-1}\sqrt{n}\alpha_{u-n}^{(r)}=0
\end{equation}
 where $\tilde{v}_{n}=(v-\tilde{\omega}_{0}n)/\eta$.

The solution reads 
\begin{equation}
c_{n}=\sum_{p=0}^{\lfloor n/2\rfloor}(-1)^{p}\left(R_{N,l}\right)^{p-n/2}\mathscr{C}_{n,p}.\label{eq:coeff}
\end{equation}
$\mathscr{C}_{n,p}$ can be regarded as a probability amplitude contribution
to the $n$-photon state from a set of corresponding qubit chain states
indexed by $p$:
\begin{equation}
\mathscr{C}_{n,p}=\frac{P_{n}}{\sqrt{n!}}\underset{\langle j_{1}\dots j_{k}\dots j_{\lfloor n/2\rfloor}\rangle}{\sum\cdots\sum}\prod_{k=1}^{p}\frac{(j_{k}+1)}{\tilde{v}_{j_{k}}\tilde{v}_{j_{k}+1}}\left[\bar{\alpha}_{u-(j_{k}+1)}^{(r)}\right]^{2}\label{eq:C_n,p}
\end{equation}
where $P_{n}=\prod_{j=0}^{n-1}\tilde{v}_{j}/\bar{\alpha}_{u-(j+1)}^{(r)}$
and $\left\langle j_{1}\dots j_{k}\dots j_{\lfloor n/2\rfloor}\right\rangle $
represents an index set of descending order $\{\forall k<l:0\leq j_{l}\leq j_{k}-2;0\leq j_{1}\leq n-2\}$.
We can see from Eq.~\eqref{eq:coeff} that the operation rules discussed
in the preceding paragraphs have made the probability amplitudes deformation-dependent,
thus $l$-dependent. This will consequently lead to a redistribution
of probabilities for different photon states. See Appendix~\ref{sub:derive_compositions}
for the derivation of these coefficients $\mathscr{C}_{n,p}$.

\begin{table}
\caption{Configurations $\left|n;r,m\right\rangle $ and probability amplitudes
$c_{n}$ for the one-excitation spectrum of an SQC with $N=4$ qubits
and spacing $l=2/3$. Each $\circ$ indicates one photon while $\uparrow$
or $\downarrow$ denotes the spin state of each qubit. Here, $\tilde{v}_{n}=(v-\tilde{\omega}_{0}n)/\eta$.~\label{tab:config}}
\vskip.1cm

\begin{tabular}{>{\centering}p{1cm}|cccc}
\hline 
$\begin{array}{c}
u=1\\
r=2
\end{array}$ & $\left\{ \begin{array}{c}
n=0\\
m=1
\end{array}\right.$ & $\left\{ \begin{array}{c}
n=1\\
m=0
\end{array}\right.$ & $\left\{ \begin{array}{c}
n=2\\
m=-1
\end{array}\right.$ & $\left\{ \begin{array}{c}
n=3\\
m=-2
\end{array}\right.$\tabularnewline
\hline 
\hline 
photon spin config. & $\begin{array}[t]{c}
-\\
\downarrow\uparrow\uparrow\uparrow\\
\uparrow\downarrow\uparrow\uparrow\\
\uparrow\uparrow\downarrow\uparrow\\
\uparrow\uparrow\uparrow\downarrow
\end{array}$ & $\begin{array}[t]{c}
\circ\\
\uparrow\uparrow\downarrow\downarrow,\downarrow\downarrow\uparrow\uparrow\\
\uparrow\downarrow\downarrow\uparrow,\downarrow\uparrow\uparrow\downarrow\\
\uparrow\downarrow\uparrow\downarrow,\downarrow\uparrow\downarrow\uparrow
\end{array}$ & $\begin{array}[t]{c}
\circ\circ\\
\uparrow\downarrow\downarrow\downarrow\\
\downarrow\uparrow\downarrow\downarrow\\
\downarrow\downarrow\uparrow\downarrow\\
\downarrow\downarrow\downarrow\uparrow
\end{array}$ & $\begin{array}[t]{c}
\circ\circ\circ\\
\downarrow\downarrow\downarrow\downarrow
\end{array}$\tabularnewline
\hline 
$c_{n}$ & 1 & $\frac{\tilde{v}_{0}}{\sqrt{6R}}$ & $\frac{\tilde{v}_{0}\tilde{v}_{1}}{6\sqrt{2}R}\negmedspace-\negthickspace\frac{1}{\sqrt{2}}$ & $\frac{\tilde{v}_{0}\tilde{v}_{1}\tilde{v}_{2}}{12\sqrt{6}R^{3/2}}\negthickspace-\negthickspace\frac{\tilde{v}_{2}+2\sqrt{6}\tilde{v}_{0}}{12\sqrt{R}}$\tabularnewline
\hline 
\end{tabular}
\end{table}

The simplest nontrivial example of  this deformation effect can be
seen in Table.~\ref{tab:config}, where we consider the one-excitation
($u=1$) spectrum of a four-qubit SQC with spacing $l=2/3$. The deformation
factor in this case is $R_{N,l}=5/8$. For a weakly coupled SQC with
$|\tilde{\omega}_{0}|\gg\eta$, the eigenenergies for the four levels
are given by (Cf. Appendix~\ref{sub:derive_4-qubit} for the derivation)
\begin{equation}
E_{\pm,\pm}=\omega_{\mathrm{q}}+\frac{3}{2}\tilde{\omega}_{0}\pm\frac{1}{2}\left[5\tilde{\omega}_{0}^{2}\pm4\tilde{\omega}_{0}\left(\tilde{\omega}_{0}^{2}+36R_{N,l}\eta^{2}\right)^{1/2}\right]^{1/2}.\label{eq:excitation_engy}
\end{equation}
For a deformation factor $R_{N,l}<1$, the splittings between these
dressed levels are suppressed. In addition, if we substitute the value
of $v$ into the coefficients $c_{n}$, we will find that $c_{1}$
is greater than that of the undeformed case, $c_{2}$ ($c_{3}$) increases
by a greater proportion than $c_{1}$ ($c_{2}$), while $c_{0}$ remains
equal to one. Therefore, the probability distribution shifts toward
the end that favors states with greater number of photons and less
degeneracy.

\section{Exciton Crossover\label{sec:crossover}}

If the spacing $l$ is large ($L_{\mathrm{q}}\gg L_{\mathrm{p}}$),
i.e., there are multiple photon wavelengths between two neighboring
qubits, we can consider the excitation that the dipole-field coupling
induces on a qubit to be localized on that qubit. Consequently, this
type of excitation emulates the Wannier exciton on an atomic lattice.
If $l$ is small ($L_{\mathrm{q}}\ll L_{\mathrm{p}}$) with a single-photon
wavelength extending over all the qubits, we can consider the excitation
on the SQC to be delocalized. This type of excitation emulates the
Frenkel's exciton model for molecular crystals. Note that when $l$
tends to either zero or infinity in Eq.~\eqref{eq:Sigma_z}, $\Sigma_{z}$
falls back to $S_{z}$ and the regular SU(2) algebra for the commutators
is obtained. Thus, it is justified that in the large-$N$ limit, the
low-energy excitation becomes bosonic for both Wannier and Frenkel
excitons.

\begin{figure}
\includegraphics[bb=137bp 220bp 515bp 555bp,clip,width=8cm]{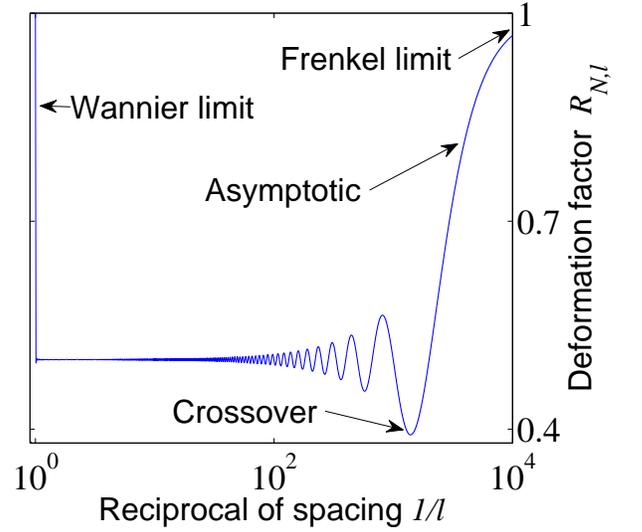}

\caption{Semi-log plot of the deformation factor $R_{N,l}$ versus $1/l$ over
one period, showing the Wannier limit on the left end and the Frenkel
limit on the right end of the horizontal axis. The number of qubits
is set to $N=1000$. \label{fig:crossover}}

\end{figure}

To determine when the emulated exciton crosses from Wannier- to Frenkel-type,
we plot in Fig.~\ref{fig:crossover} the deformation factor versus
the reciprocal of the spacing over $10^{-4}<l<1$, where we can see
the asymmetry between the left edge ($1/l\to0$) for the Wannier limit
and the right edge ($1/l\to\infty$) for the Frenkel limit. Setting
$\mathrm{d}R_{N,l}/\mathrm{d}l=0$, we obtain the trigonometric equation
\begin{equation}
\tan[(2N-1)\pi l]=(2N-1)\tan(\pi l).\label{eq:crsvr_eq}
\end{equation}
Transforming Eq.~\eqref{eq:crsvr_eq} to $U_{2N-1}(\cos\pi l)=2N\cdot T_{2N-1}(\cos\pi l)$,
where $T_{2N-1}$ ($U_{2N-1}$) is the Chebyshev polynomial of the
first (second) kind, we can observe that it is a $(2N-1)$-th order
polynomial equation. Hence, the curve has $(2N-1)$ local extrema
in exactly $(N-1)$ oscillations from the Wannier end to the asymptotic
Frenkel end. Between these two limits, the excitation has various
degrees of deformation and the crossover is continuous. We can regard
the crossover point to be the absolute minimum before the deformation
factor asymptotically approaches one. This point approaches 0 when
$N\to\infty$. For the case illustrated in Fig.~\ref{fig:crossover}
with $N=1000$, a numerical estimation gives the crossover at $l=7.16\times10^{-4}$,
or a length of 2800 spins per photon wavelength.

\section{Conclusion}

We have studied the inhomogeneous coupling between a SQC and a superconducting
coplanar resonator, which leads to a set of deformation-dependent
operation rules of spin momentum. The modified rules correspond to
tighter energy spacings and a shift of the probability distribution
of spin levels. The inhomogeneous coupling also gives rise to different
types (Frenkel and Wannier) of collective excitations on the SQC and
the crossover between these types is determined by a polynomial equation
of the qubit spacing $l$.

FN acknowledges partial support from LPS, NSA, ARO, DARPA, AFOSR,
NSF grant No.~0726909, JSPSRFBR contract No.~09-02-92114, MEXT Kakenhi
on Quantum Cybernetics, and Funding Program for Innovative R\&D on
S\&T. YXL acknowledges support from the NNSFC grant No.~10975080
and 61025022.

\appendix

\section{Derivation of the deformation factor}

\subsection{Commutation relation\label{sub:comm_rel}}

The commutator of the ladder operators can be computed as follows
\begin{eqnarray*}
[S_{+},S_{-}] & = & \sum_{j,k=0}^{N-1}\cos(j\pi l)\cos(k\pi l)[\sigma_{j,+},\sigma_{k,-}]\\
 & = & 2\sum_{j=0}^{N-1}\cos^{2}(j\pi l)\sigma_{j,z}\\
 & = & S_{z}+\sum_{j=0}^{N-1}\cos(2j\pi l)\sigma_{j,z}.
\end{eqnarray*}
To use a consistent notation, we write the R.H.S. as $2\Sigma_{z}$
and extract $S_{z}$ from the second term 
\begin{eqnarray*}
2\Sigma_{z} & = & 2S_{z}+\sum_{j=0}^{N-1}\left[\cos(2j\pi l)-\cos(2j\pi)\right]\sigma_{j,z}\\
 & = & 2S_{z}+2\sum_{j=0}^{N-1}\sin\left[j\pi(1+l)\right]\sin\left[j\pi(1-l)\right]\sigma_{j,z}
\end{eqnarray*}
which gives Eq.~\eqref{eq:Sigma_z}.

We can also check that other commutation relations are preserved
\begin{eqnarray*}
[S_{z},S_{\pm}] & = & \sum_{j,k=0}^{N-1}\left[\sigma_{k,z},\sigma_{j,\pm}\cos(j\pi l)\right]\\
 & = & \sum_{j=0}^{N}\pm\sigma_{j,\pm}\cos(j\pi l)\\
 & = & \pm S_{\pm},
\end{eqnarray*}
from which we conclude that the newly defined operators $\left\{ S_{z},S_{+},S_{-}\right\} $
form a Polychronakos-Rocek type of deformed SU(2) algebra.

\subsection{Deformation factor\label{sub:defm_factor}}

First, we recognize that each qubit $\sigma_{j,z}$ has two orthonormal
basis vectors $\left\{ \left|e_{j}\right\rangle ,\left|g_{j}\right\rangle \right\} $
for a fixed relative coordinate $j$ and hence $\left\{ \left|\epsilon_{k}\right\rangle :\left|\epsilon_{k}\right\rangle \in\left\{ \left|e_{j}\right\rangle ,\left|g_{j}\right\rangle ,j=0,\dots,N-1\right\} \right\} $
forms an orthonormal basis for the Hilbert space $\mathcal{H}$ that
spans all the qubits on the chain. Then $S_{z}$ for the original
spin angular momentum and $\Sigma_{z}$ for that of the inhomogeneous
SQC become operators on this Hilbert space $\mathcal{H}$. Since the
sinusoidal functions $\cos(j\pi l)$ are bounded, we can define the
Hilbert-Schmidt inner product as
\begin{eqnarray*}
\left\langle \Sigma_{z},S_{z}\right\rangle  & = & \mathrm{tr}\left(\Sigma_{z}^{\ast}S_{z}\right)=\sum_{k}^{2N}\left\langle \epsilon_{k}\right|\Sigma_{z}\cdot S_{z}\left|\epsilon_{k}\right\rangle \\
 & = & \sum_{k}^{2N}\left\langle \epsilon_{k}\right|\sum_{j,l=0}^{N-1}\cos^{2}(j\pi l)\sigma_{j,z}\cdot\sigma_{l,z}\left|\epsilon_{k}\right\rangle \\
 & = & \sum_{k}^{2N}\left\langle \epsilon_{k}\right|\cos^{2}(\lfloor k/2\rfloor\pi l)\sigma_{\lfloor k/2\rfloor,z}\cdot\sigma_{\lfloor k/2\rfloor,z}\left|\epsilon_{k}\right\rangle \\
 & = & \sum_{k}^{2N}\cos^{2}(\lfloor k/2\rfloor\pi l)\\
 & = & \sum_{j}^{N}2\cos^{2}(j\pi l)
\end{eqnarray*}
which equals to Eq.~\eqref{eq:HS_norm}. We can then write the approximation
of $\Sigma_{z}$ as a Schmidt projection on $S_{z}$
\begin{eqnarray*}
\Sigma_{z} & \approx & \frac{\left\langle \Sigma_{z},S_{z}\right\rangle }{\left\langle S_{z},S_{z}\right\rangle }S_{z}\\
 & = & \frac{N+1+\sum_{j}\cos(2j\pi l)}{2(N+1)}S_{z}\\
 & = & \left[\frac{1}{2}+\frac{1}{2N}\sum_{j=0}^{N-1}\cos(2j\pi l)\right]S_{z}\\
 & = & R_{N,l}S_{z}
\end{eqnarray*}
where $R_{N,l}$ denotes the deformation factor. Its expression can
be further simplified to
\begin{eqnarray*}
R_{N,l} & = & \frac{1}{4N}\left[2N+1-\cos(2N\pi l)+\frac{\sin(2\pi l)\sin(2N\pi l)}{1-\cos(2\pi l)}\right]\\
 & = & \frac{1}{4N}\left[2N+1-\cos(2N\pi l)+\frac{\cos(\pi l)\sin(2N\pi l)}{\sin(\pi l)}\right]\\
 & = & \frac{1}{4N}\left[2N+1+\frac{\sin[(2N-1)\pi l]}{\sin(\pi l)}\right]
\end{eqnarray*}
where the first line is derived by comparing the real parts in a summation
of exponentials.

\subsection{Casimir operator\label{sub:csmir_op}}

The Casimir operator for the algebra is
\[
C=S_{-}S_{+}+h(S_{z})
\]
where the second term satisfies a recursive relation~\cite{delbecq93}
\[
h(S_{z})-h(S_{z}-1)=2R_{N,l}S_{z}.
\]
This relation leads to a solution composed of Bernoulli polynomials
\begin{eqnarray*}
h(S_{z}) & = & R_{N,l}\left(B_{2}(-S_{z})-B_{2}\right)\\
 & = & R_{N,l}\left(S_{z}^{2}+S_{z}\right)
\end{eqnarray*}
where $B_{2}(-S_{z})$ is the second-order Bernoulli polynomial with
the operator $S_{z}$ as variable and $B_{2}$ is the second Bernoulli
number. The Casimir operator becomes then
\begin{eqnarray*}
C & = & S_{-}S_{+}+R_{N,l}\left(S_{z}^{2}+S_{z}\right)\\
 & = & \frac{1}{2}(S_{+}S_{-}+S_{-}S_{+})+R_{N,l}S_{z}^{2}
\end{eqnarray*}
which equals to Eq.~\eqref{eq:csmir_op} and represents a deformed
total spin operator $S^{2}$.

\section{Deriving operation rules\label{sec:derive_op_rules}}

Assume the eigenstate of the $z$-spin momentum operator $S_{z}$
to be $\left|r,m\right\rangle $, that is, $r(r+1)$ denotes the total
spin number and $m$ the magnetic moment, for which Eqs.~\eqref{eq:S_z}-\eqref{eq:S^2}
are satisfied. Further, assume $\alpha_{m}^{(r)}$ to be the coefficients
when the ladder operators are applied to the state vectors as in Eqs.~\eqref{eq:S_+}-\eqref{eq:S_-},
which is indexed by $r$ and $m$. 

By applying the vector $\left|r,m\right\rangle $ to the commutation
relation Eq.~\eqref{eq:defm_comm_rel}, we find
\begin{eqnarray*}
\left\langle r,m\right|\left[S_{+},S_{-}\right]\left|r,m\right\rangle  & = & \left\langle r,m\right|S_{+}S_{-}-S_{-}S_{+}\left|r,m\right\rangle \\
 & = & (\alpha_{m-1}^{(r)})^{2}-(\alpha_{m}^{(r)})^{2}\\
 & = & \left\langle r,m\left|2R_{N,l}S_{z}\right|r,m\right\rangle \\
 & = & 2R_{N,l}m.
\end{eqnarray*}
Selecting the second and the fourth line, we arrive at a difference
equation of $m$:
\[
(\alpha_{m}^{(r)})^{2}-(\alpha_{m-1}^{(r)})^{2}=-2mR_{N,l}.
\]
To solve the equation, we list out the iterations until the last entry
where $\alpha_{-r-1}^{(r)}=0$ since $-r$ is the minimum value $m$
can take as the magnetic moment 
\begin{eqnarray*}
(\alpha_{m}^{(r)})^{2}-(\alpha_{m-1}^{(r)})^{2} & = & -2R_{N,l}m\\
(\alpha_{m-1}^{(r)})^{2}-(\alpha_{m-2}^{(r)})^{2} & = & -2R_{N,l}(m-1)\\
\vdots & \vdots & \vdots\\
(\alpha_{-r}^{(r)})^{2}-(\alpha_{-r-1}^{(r)})^{2} & = & -2R_{N,l}(-r).
\end{eqnarray*}
Summing up all the iterations above, we have
\begin{eqnarray}
(\alpha_{m}^{(r)})^{2} & = & -2R_{N,l}\sum_{j=0}^{m+r}(m-j)\nonumber \\
 & = & -R_{N,l}(m-r)(m+r+1),\label{eq:alpha^2}
\end{eqnarray}
which gives Eq.~\eqref{eq:op_rule}. We can verify this result by
summing up instead of summing down, i.e., with the condition $\alpha_{r}^{(r)}=0$
and the iterations 
\begin{eqnarray*}
(\alpha_{m}^{(r)})^{2}-(\alpha_{m+1}^{(r)})^{2} & = & 2R_{N,l}(m+1)\\
(\alpha_{m+1}^{(r)})^{2}-(\alpha_{m+2}^{(r)})^{2} & = & 2R_{N,l}(m+2)\\
\vdots & \vdots & \vdots\\
(\alpha_{r-1}^{(r)})^{2}-(\alpha_{r}^{(r)})^{2} & = & 2R_{N,l}r
\end{eqnarray*}
we have, after adding them up,
\begin{eqnarray*}
(\alpha_{m}^{(r)})^{2} & = & 2R_{N,l}\sum_{j=0}^{r-m}(m+j)\\
 & = & R_{N,l}(r-m+1)(m+r),
\end{eqnarray*}
which is the same as Eq.~\eqref{eq:alpha^2}.

\section{Deriving the excitation spectrum of the superconducting qubit chain}

\subsection{State vector compositions for a general $N$-qubit superconducting
qubit chain\label{sub:derive_compositions}}

The form into which the system Hamiltonian is split as in Eqs.~\eqref{eq:H_0}-\eqref{eq:H_1}
ensures that $[H_{0},H_{1}]=0$. The commutation of these two parts
implies that we can find simultaneous eigenvectors for $H_{0}$ and
$H_{1}$.

First, for an eigenvector $\left|n;r,m\right\rangle $ (or written
as $\left|u,r\right\rangle $) of $H_{0}$, we have
\[
H_{0}\left|n;r,m\right\rangle =H_{0}\left|u,r\right\rangle =\omega_{q}(m+n)=\omega_{q}u
\]
where $\{u,n,m\}$ assumes meanings as described in Sec.~\ref{sec:spectrum}.
Note that the eigenvalue $\omega_{q}u$ is degenerate, for different
combinations of $m$ and $n$ that add up to the same $u$. Therefore
the eigenstate of $H_{0}$ can be written as a superposition
\begin{eqnarray}
\left|u,r\right\rangle  & = & \sum_{n,m}c_{n}\left|n;r,m\right\rangle \delta(u-n-m)\nonumber \\
 & = & \sum_{n}c_{n}\left|n;r,u-n\right\rangle \Delta\label{eq:eigenvec}
\end{eqnarray}
where $\Delta$ is a range delta function
\[
\Delta=\begin{cases}
1, & -r\leq u-n\leq r\\
0, & \mathrm{otherwise}
\end{cases}
\]
since we have to ensure the state vectors satisfy the addition rules
of angular momentum.

Our next step is to find those of Eq.~\eqref{eq:eigenvec} that are
also simultaneous eigenvectors of $H_{1}$. With the modified operation
rule Eq.~\eqref{eq:op_rule} and setting $m=u-n$, we can apply $H_{1}$
to the expression and, after reshuffling the terms in the summation
such that vectors with the same total excitation number are grouped
together, we find 
\begin{multline}
H_{1}\left|u,r\right\rangle =\sum_{n}\biggl\{ c_{n}\tilde{\omega}_{0}n+c_{n+1}\eta\sqrt{(n+1)}\alpha_{u-n-1}^{(r)}\\
+c_{n-1}\eta\sqrt{n}\alpha_{u-n}^{(r)}\biggr\}\left|n;r,u-n\right\rangle \Delta.\label{eq:H1_vec_1}
\end{multline}
Since $\nu$ is the eigenvalue of $H_{1}$, we have
\begin{equation}
H_{1}\left|u,r\right\rangle =v\left|u,r\right\rangle =\sum_{n}c_{n}v\left|n;r,u-n\right\rangle \Delta\label{eq:H1_vec_2}
\end{equation}
Then comparing Eq.~\eqref{eq:H1_vec_1} with Eq.~\eqref{eq:H1_vec_2},
we deduce a difference equation
\[
c_{n+1}\eta\sqrt{(n+1)}\alpha_{u-(n+1)}^{(r)}-c_{n}\tilde{v}_{n}+c_{n-1}\eta\sqrt{n}\alpha_{u-n}^{(r)}=0
\]
where $\tilde{v}_{n}=(v-\tilde{\omega}_{0}n)/\eta$ and the initial
conditions are
\begin{eqnarray*}
c_{-1} & = & 0,\\
c_{u+r+1} & = & 0.
\end{eqnarray*}
In addition, from the definition of the $\Delta$ function, $n\leq u+r$.

Now write
\begin{equation}
c_{n}=\frac{C_{n}}{\sqrt{n!}\prod_{j=1}^{n}\alpha_{u-j}^{(r)}}\label{eq:coe_tfm}
\end{equation}
and we have a simplified difference equation
\[
C_{n+1}-C_{n}\tilde{v}_{n}+C_{n-1}n\left(\alpha_{u-n}^{(r)}\right)^{2}=0.
\]
To find the solution, we multiply each equation starting with $C_{j}$
by $\prod_{k=j}^{n}\tilde{v}_{n}$ 
\begin{eqnarray*}
C_{n}\tilde{v}_{n}-C_{n-1}\tilde{v}_{n}\tilde{v}_{n-1}+(n-1)\tilde{v}_{n}C_{n-2}\left[\alpha_{u-(n-1)}^{(r)}\right]^{2} & = & 0\\
\vdots & \vdots & \vdots\\
C_{2}\prod_{j=2}^{n-1}\tilde{v}_{j}-C_{1}\prod_{j=1}^{n-1}\tilde{v}_{j}+C_{0}\left[\alpha_{u-1}^{(r)}\right]^{2}\prod_{j=2}^{n-1}\tilde{v}_{j} & = & 0\\
C_{1}\prod_{j=1}^{n-1}\tilde{v}_{j}-C_{0}\prod_{j=0}^{n-1}\tilde{v}_{j} & = & 0
\end{eqnarray*}
Then with the terminating conditions $C_{1}=C_{0}\tilde{v}_{0}$ and
$C_{0}=1$, we can sum up the equations to eliminate the middle terms
and obtain
\[
C_{n}-Q_{0,n-1}+\sum_{j=0}^{n-2}(j+1)C_{j}\alpha_{u-(j+1)}^{2}Q_{j+2,n-1}=0
\]
where we use a shorthand notation
\[
Q_{0,n-1}=\prod_{j=0}^{n-1}\tilde{v}_{j}.
\]

To find the analytical expression for $C_{n}$, we recursively expand
the factor $C_{j}$ 
\begin{eqnarray*}
C_{n} & = & Q_{0,n-1}-\sum_{j=0}^{n-2}(j+1)C_{j}\alpha_{u-(j+1)}^{2}Q_{j+2,n-1}\\
 & = & Q_{0,n-1}-\sum_{j=0}^{n-2}Q_{0,j-1}Q_{j+2,n-1}(j+1)\alpha_{u-(j+1)}^{2}\\
 &  & +\sum_{j=0}^{n-2}\sum_{k=0}^{j-2}Q_{j+2,n-1}Q_{0,k-1}Q_{k+2,j-1}\\
 &  & \times\left[(j+1)\alpha_{u-(j+1)}^{2}\right]\left[(k+1)\alpha_{u-(k+1)}^{2}\right]-\cdots
\end{eqnarray*}
By observing that $Q_{0,j-1}Q_{j+2,n-1}=Q_{0,n-1}/\tilde{v}_{j}\tilde{v}_{j+1}$
and so on for each pair of $Q$'s in the terms of each recursive expansion,
we can recursively factorize out $Q_{0,n-1}$ and arrive at
\begin{multline*}
C_{n}=Q_{0,n-1}\sum_{p=0}^{\lfloor n/2\rfloor}(-1)^{p}\underset{\langle j_{1}\dots j_{k}\dots j_{\lfloor n/2\rfloor}\rangle}{\sum\cdots\sum}\\
\prod_{k=1}^{p}\frac{(j_{k}+1)}{\tilde{v}_{j_{k}}\tilde{v}_{j_{k}+1}}\left[\alpha_{u-(j_{k}+1)}^{(r)}\right]^{2}
\end{multline*}
where $\left\langle j_{1}\dots j_{k}\dots j_{\lfloor n/2\rfloor}\right\rangle $
is the index set described in Sec.~\ref{sec:spectrum}. Finally,
substituting the above expression back to the transformation Eq.~\eqref{eq:coe_tfm},
we can obtain the coefficients of the excitation eigenvector as in
Eq.~\eqref{eq:coeff}.

\subsection{One-excitation spectrum for a 4-qubit superconducting qubit chain\label{sub:derive_4-qubit}}

The state vector for the one-excitation 4-qubit SQC ($u=1$, $r=2$,
and $m\in\{-2,-1,0,1\}$) can be written as
\[
\left|u,r\right\rangle =c_{0}\left|0;2,1\right\rangle +c_{1}\left|1;2,0\right\rangle +c_{2}\left|2;2,-1\right\rangle +c_{3}\left|3;2,-2\right\rangle .
\]
If we assume $c_{0}=1$ as a common factor, the rest three coefficients
can be written as
\begin{eqnarray*}
c_{1} & = & R^{-1/2}\mathscr{C}_{1,0}\\
c_{2} & = & R^{-1}\mathscr{C}_{2,0}-\mathscr{C}_{2,1}\\
c_{3} & = & R^{-3/2}\mathscr{C}_{3,0}-R^{-1/2}\mathscr{C}_{3,1}.
\end{eqnarray*}
After plugging in the expression according to Eq.~\eqref{eq:C_n,p},
we obtain the expressions shown in Table.~\ref{tab:config}.

To find $v$, and hence the excitation energy, consider the difference
equations
\begin{eqnarray*}
C_{4}-C_{3}\tilde{v}_{3}+3C_{2}(\alpha_{-2}^{(2)})^{2} & = & 0\\
C_{3}-C_{2}\tilde{v}_{2}+2C_{1}(\alpha_{-1}^{(2)})^{2} & = & 0\\
C_{2}-C_{1}\tilde{v}_{1}+C_{0}(\alpha_{0}^{(2)})^{2} & = & 0.
\end{eqnarray*}
Since $C_{4}=0$ and $C_{1}=\tilde{v}_{0}C_{0}$, we derive from the
last equation 
\[
C_{2}=C_{0}\left[\tilde{v}_{0}\tilde{v}_{1}-(\alpha_{0}^{(2)})^{2}\right]
\]
and from the first equation
\[
C_{3}=3C_{0}\left[\tilde{v}_{0}\tilde{v}_{1}+(\alpha_{0}^{(2)})^{2}\right]\frac{(\alpha_{-2}^{(2)})^{2}}{\tilde{v}_{3}}.
\]
Substitute these expressions into the second equation and with Eq.~\eqref{eq:op_rule},
we have
\begin{multline*}
12(\tilde{v}_{0}\tilde{v}_{1}+6R_{N,l})R_{N,l}-(\tilde{v}_{0}\tilde{v}_{1}-6R_{N,l})\tilde{v}_{2}\tilde{v}_{3}\\
+12R_{N,l}\tilde{v}_{0}\tilde{v}_{3}=0.
\end{multline*}
Expanding the $\tilde{v}$, we arrive at a fourth-order polynomial
equation
\begin{multline*}
v^{4}-6\tilde{\omega}_{0}v^{3}+\left[11\tilde{\omega}_{0}^{2}-30R_{N,l}\eta^{2}\right]v^{2}\\
-6\left[\tilde{\omega}_{0}^{3}-13\tilde{\omega}_{0}R_{N,l}\eta^{2}\right]v\\
-36R_{N,l}\eta^{2}\left[\tilde{\omega}_{0}^{2}+2R_{N,l}\eta^{2}\right]=0.
\end{multline*}
If we consider the case with $\tilde{\omega}_{0}=0$ (the conventional
TC-model case), we have 
\[
v^{4}-30R_{N,l}\eta^{2}v^{2}-72R_{N,l}^{2}\eta^{4}=0
\]
and the roots are
\[
v=\pm\sqrt{(15+3\sqrt{33})R_{N,l}}\eta
\]
On the other hand, if we consider a weak coupling case $\tilde{\omega}_{0}\gg\eta$,
then the equation becomes
\[
v^{4}-6\tilde{\omega}_{0}v^{3}+11\tilde{\omega}_{0}^{2}v^{2}-6\tilde{\omega}_{0}^{3}v-36R_{N,l}\eta^{2}\tilde{\omega}_{0}^{2}=0
\]
and the solutions are
\[
v=\frac{3}{2}\tilde{\omega}_{0}\pm\frac{1}{2}\sqrt{5\tilde{\omega}_{0}^{2}\pm4\tilde{\omega}_{0}\sqrt{\tilde{\omega}_{0}^{2}+36R_{N,l}\eta^{2}}}.
\]
Hence, the excitation energy can be written as in Eq.~\eqref{eq:excitation_engy}.

Now the coefficient $c_{1}$ for the highest eigenenergy state is,
since usually $\tilde{\omega_{0}}<0$,
\[
c_{1}=\frac{v}{\sqrt{6R_{N,l}}\eta}=\frac{3}{2}\Omega+\frac{1}{2}\sqrt{5\Omega^{2}-4\Omega\sqrt{\Omega^{2}+6}}
\]
where $\Omega=\tilde{\omega}_{0}/(\eta\sqrt{6R_{N,l}})$. Since $R_{N,l}\leq1$,
\[
|\Omega|\geq\left|\frac{\tilde{\omega}_{0}}{\sqrt{6}\eta}\right|
\]
which means that the deformation leads to a larger coefficient $c_{1}$
than that of the undeformed case. For $c_{2}$, we have
\[
c_{2}=\frac{1}{\sqrt{2}}\left(c_{1}^{2}-\Omega c_{1}+1\right)>\frac{1}{\sqrt{2}}(c_{1}+1)^{2}
\]
since $\tilde{\omega}_{0}\gg\eta$ and so $|\Omega|\gg2$. This means
that $c_{2}$ increases by a greater proportion than $c_{1}$ due
to the deformation. Similarly $c_{3}$ increasing by an even larger
factor. The change in the coefficients shows that the a larger deformation
favors the states with larger number of photons and a more ordered
spin-chain.

\end{document}